%A SET OF MACROS FOR WRITING PAPERS.
%
%\today GIVES TODAY'S DATE.
%
\def\today{\ifcase\month\or January\or February\or March\or April\or May\or
June\or July\or August\or September\or October\or November\or December\fi
\space\number\day, \number\year}
%
%\note{footnote} GIVES SEQUENTIALLY NUMBERED FOOTNOTES.
%
\newcount\notenumber

\def\note{\global\advance\notenumber by 1 \footnote{$^{\the\notenumber}$}}
%
%\numbereq SEQUENTIALLY NUMBERS EQUATIONS ON THE RIGHT (number)
%
\newif\ifsectionnumbering
\newcount\eqnumber
\def\cleareqnumber{\eqnumber=0}
\def\numbereq{\global\advance\eqnumber by 1
\ifsectionnumbering\eqno(\the\secnumber.\the\eqnumber)\else\eqno
(\the\eqnumber)\fi}
\def\eqalinno{{\global\advance\eqnumber by 1}
\ifsectionnumbering(\the\secnumber.\the\eqnumber)\else(\the\eqnumber)\fi}
\def\name#1{\ifsectionnumbering\xdef#1{\the\secnumber.\the\eqnumber}
\else\xdef#1{\the\eqnumber}\fi}
\def\nosectionnumbering{\sectionnumberingfalse}
\sectionnumberingtrue
%
%\ref{\name} GIVES SEQUENTIALLY NUMBERED
%REFERENCES [number], AND ASSIGNS
%THAT NUMBER TO A MACRO \name AND WRITES REF. TO FILE 1.
%
\newcount\refnumber

\immediate\openout1=refs.tex
\immediate\write1{\noexpand\frenchspacing}
\immediate\write1{\parskip=0pt}
\def\ref#1#2{\global\advance\refnumber by 1%
[\the\refnumber]\xdef#1{\the\refnumber}%
\immediate\write1{\noexpand\item{[#1]}#2}}
\def\tie{\noexpand~}

%
% NEW SECTION: \newsection The Method. (terminate with a .)
%
\font\twelvebf=cmbx10 scaled \magstep1
\newcount\secnumber

\def\newsection#1.{\ifsectionnumbering\cleareqnumber\else\fi%
	\global\advance\secnumber by 1%
	\bigbreak\bigskip\par%
	\line{\twelvebf \the\secnumber. #1.\hfil}\nobreak\medskip\par}
%
%
%\Box GIVES WAVE OPERATOR, OR LAPLACIAN
%
\def \sqr#1#2{{\vcenter{\vbox{\hrule height.#2pt
	\hbox{\vrule width.#2pt height#1pt \kern#1pt
		\vrule width.#2pt}
		\hrule height.#2pt}}}}
\def\Box{{\mathchoice\sqr54\sqr54\sqr33\sqr23}\,}
%
%
%\twocolumns GIVES TWO-COLUMN OUTPUT
%
\newdimen\fullhsize
\def\fiddle{\fullhsize=6.5truein \hsize=3.2truein}
\def\fullline{\hbox to\fullhsize}
\def\mkhdline{\vbox to 0pt{\vskip-22.5pt
	\fullline{\vbox to8.5pt{}\the\headline}\vss}\nointerlineskip}
\def\mkftline{\baselineskip=24pt\fullline{\the\footline}}
\let\lr=L \newbox\leftcolumn
\def\twocolumns{\fiddle
	\output={\if L\lr \global\setbox\leftcolumn=\columnbox
		\global\let\lr=R \else \doubleformat \global\let\lr=L\fi
		\ifnum\outputpenalty>-20000 \else\dosupereject\fi}}
\def\doubleformat{\shipout\vbox{\mkhdline
		\fullline{\box\leftcolumn\hfil\columnbox}
		\mkftline} \advancepageno}
\def\columnbox{\leftline{\pagebody}}
\nosectionnumbering
\magnification=1200
\def\pr#1 {Phys. Rev. {\bf D#1\tie }}
\def\pe#1 {Phys. Rev. {\bf #1\tie}}
\def\pre#1 {Phys. Rep. {\bf #1\tie}}
\def\pl#1 {Phys. Lett. {\bf #1B\tie }}
\def\prl#1 {Phys. Rev. Lett. {\bf #1\tie }}
\def\np#1 {Nucl. Phys. {\bf B#1\tie }}
\def\ap#1 {Ann. Phys. (NY) {\bf #1\tie }}
\def\cmp#1 {Commun. Math. Phys. {\bf #1\tie }}
\def\imp#1 {Int. Jour. Mod. Phys. {\bf A#1\tie }}
\def\mpl#1 {Mod. Phys. Lett. {\bf A#1\tie}}
\def\zp#1 {Z. Phys. {\bf C#1\tie}}
\def\amp#1 {Adv. Theor. Math. Phys. {\bf#1\tie}}
\def\tie{\noexpand~}

\def\s{(\sigma)}

\def\sp{(\sigma ')}

\def\Tb{\overline T}
\def\d{\delta(\sigma-\sigma ')}
\def\dpr{\delta'(\sigma-\sigma ')}
\def\dppr{\delta''(\sigma-\sigma ')}
\def\dpppr{\delta'''(\sigma-\sigma ')}

\def\ints{\int d\sigma\,}

\def\dx{\partial X}

\parskip=15pt plus 4pt minus 3pt
\headline{\ifnum \pageno>1\it\hfil Superconformal Deformations
and Spacetime	$\ldots$\else \hfil\fi}
\font\title=cmbx10 scaled\magstep1
\font\tit=cmti10 scaled\magstep1
\footline{\ifnum \pageno>1 \hfil \folio \hfil \else
\hfil\fi}
\raggedbottom
\rightline{\vbox{\hbox{NYU-TH/99-02-01}\hbox{hep-th/yymmnn}}}
\vfill
\centerline{\title SUPERCONFORMAL DEFORMATIONS AND}
\vskip 20pt
\centerline{\title SPACE-TIME SYMMETRIES}
\vfill
{\centerline{\title
Ioannis Giannakis \footnote{$^{\dag}$}
{\rm e-mail:ig231@scires.nyu.edu}}
}
\medskip
\centerline{\tit Department of Physics, New York
University}
\centerline{\tit 4 Washington Pl., New York, NY
10003.}
\vfill
\centerline{\title Abstract}
\bigskip
{\narrower\narrower
In this paper we present a method of deforming
to first order the stress-tensor and the supercurrent of the
superstring corresponding to turning on NS-NS
bosonic fields. Furthermore
we discuss the difficulties associated with turning on spacetime
fermions and R-R bosons. We also
derive the gauge symmetries of the massless spacetime fields.
\par}

\vfill\vfill\break

\newsection Introduction.

Superconformal field theories ( with appropriate central charge ) are
solutions to the classical equations of motion of string theory.
Thus by studying infinitesimal deformations which preserve this
superconformal structure \ref{\eaw}{J. Freericks and M. Halpern, 
\ap188 258 (1988).}, we are investigating the linearised
classical equation of motion about the corresponding solution.
This is an interesting problem in its own right, but it also provides us with
insights into the symmetry structure of string theory since a spacetime
symmetry transformation is a particular deformation.
To study symmetries, we seek transformations of the
space-time fields that take one solution of the classical
equations of motion to another that is physically equivalent.
Since, ``Solutions of the classical equations
of motion," are two-dimensional superconformal
field theories,
we are thus interested in {\it isomorphic superconformal
field theories}.

\newsection Superconformal Deformations.
 
We shall work in a Hamiltonian formalism.
Any quantum mechanical theory (including a superconformal
field theory) is defined
by three elements: i)~an algebra
of operator valued distributions,
usually called superfields, $\cal A$ (determined by the degrees of
freedom of the theory
and their equal-time commutation relations),  ii)~a
representation of that algebra
and iii)~two distinguished elements of $\cal A$, 
${\cal T}(\sigma, \theta)=T_F(\sigma)+{\theta}T(\sigma)$
and ${\overline {\cal T}}(\sigma, \theta)=
{\overline T}_F(\sigma)+{\overline{\theta}}{\overline T}(\sigma)$.
Superconformal operator algebras include also spin fields.
In terms of these fields the Hamiltonian $H$ and the
generator of translations $P$
are given by
$$
H=\int d\sigma (T\s + \Tb\s), \qquad
P=\int d\sigma (T\s - \Tb\s).
\numbereq\name{\eqaws}
$$
Also the components $T_F(\sigma)$ and $T(\sigma)$
must satisfy two mutually commuting copies of the
SuperVirasoro algebra ( we have omitted one copy of the SuperVirasoro
for simplicity ):
$$
\eqalignno{[T\s, T\sp]&={- i c \over 24\pi}\dpppr
+2 i T\sp\dpr -  i T'\sp\d&{\global\advance\eqnumber by 1}
(\the\secnumber.\the\eqnumber a)\name{\eqvir}\cr
\{ T\s, T_F\sp \}&=-{1\over 2{\sqrt 2}}T\sp\d+{c\over 24{\sqrt 2}{\pi}}
\dppr&(\the\secnumber.\the\eqnumber
b)\cr
[T\s, T_F\sp]&={3i\over 2}T_F\sp\dpr-
iT'_F\sp\d.&(\the\secnumber.\the\eqnumber c)\cr}
$$
The existence of the SuperVirasoro algebra means that the states of
the theory can be organised into modules of that algebra.
If the Hamiltonian, Eq. (\eqaws), is bounded below, these are
highest weight representations. The highest weight states of these
representations are created by the other important
operators in the theory, the {\it superconformal primary fields}
which are constructed from elementary fields and momenta and are defined as 
any pair of fermionic $\Phi_F(\sigma)$ and bosonic
$\Phi_B(\sigma)$ fields that satisfy
$$
\eqalignno{[T\s, {\Phi_F}\sp]&=
id {\Phi_F}\sp\dpr - {i\over {\sqrt 2}}{\partial}
{\Phi_F}\sp\d&{\global\advance\eqnumber by 1}
(\the\secnumber.\the\eqnumber a)\name{\eqvir}\cr
[T\s, {\Phi_B}\sp]&=
i(d+{1\over 2}){\Phi_B}\sp\dpr - {i\over {\sqrt 2}}{\partial}
{\Phi_B}\sp\d&
(\the\secnumber.\the\eqnumber b)\name{\eqvir}\cr
\{ T_F\s, {\Phi_F}\sp \}&=-{1\over 2{\sqrt 2}}{\Phi_B}\sp\d
&(\the\secnumber.\the\eqnumber c)\cr
[T_F\s, {\Phi_B}\sp]&=
id{\Phi_F}\sp\dpr - {i\over 2{\sqrt 2}}{\partial}
{\Phi_F}\sp\d&
(\the\secnumber.\the\eqnumber d)\name{\eqvir}\cr}
$$

Note that for the same $\cal A$ we may
have many choices
of Hamiltonian, so that $\cal A$ should more properly be
associated with a
deformation class of superconformal theories than with one
particular theory. The superfields do not exhaust the set of
all operators in the algebra $\cal A$. On the complex plane for example
the fermionic components of the superfields as two-dimensional
spinors are double-valued fields,
$\Phi_{F}(e^{2{\pi}i}z)=\pm\Phi_{F}(z)$.
This implies that the operator
algebra $\cal A$ encompasess spin fields $S^{\alpha}(z)$
whose presence modifies the boundary conditions of the fermionic
components of the superfields.
As a result the operator algebra is not local since the
OPE of the fermionic component of a superfield
$\Phi_F(z)$ with a spin field $S^{\alpha}(w)$ includes half integral powers
of ${1\over (z-w)}$. Locality appears to be essential in order to have a well
defined string theory.
We can either restrict ourselves to one
of the two boundary conditions for the fermionic components of
the superfields or include both boundary conditions
(NS and R) but by eliminating half of the operators of each we regain
a local operator algebra. This will involve the projection
(GSO projection ) of the
non-local operator algebra $\cal A$ onto a local one ${\cal A}_1$.

String theory requires the full structure of the SuperVirasoro algebras
in order to decouple negative norm states from physical processes.

We are interested in not just one, but a family of superconformal
field theories parametrized by the values of the spacetime fields.
Changing these spacetime fields changes the superconformal field
theory but preserves the SuperVirasoro algebra (including the value
of the central charge).
We will discuss deformations of the local operator algebra
${\cal A}_1$ which involve a change in our
choice of $T, T_{F}, {\overline T}$ and ${\overline T_{F}}$ in a way that preserves
the SuperVirasoro algebra. Thus under deformations
$$
T\s \rightarrow T\s+{\delta}T\s \qquad T_{F}\s \rightarrow T_{F}\s+
{\delta}T_{F}\s
\numbereq\name{\eqwas}
$$
and the preservation of the SuperVirasoro algebra, to first order
in variations, we require
$$
\eqalign{
&[{\delta}T\s, T\sp]+[T\s, {\delta}T\sp]=
2 i{\delta}T\sp\dpr -  i{\delta}T'\sp\d \cr
&\{ {\delta}T_F\s, T_F\sp \}+\{ T_F\s, {\delta}T_F\sp \}=-{1\over 2{\sqrt 2}}
{\delta}T\sp\d \cr
&[{\delta}T\s, T_F\sp]+[T\s, {\delta}T_F\sp]={3i\over 2}{\delta}T_F\sp\dpr-
i{\delta}T'_F\sp\d. \cr}
\numbereq\name{\eqana}
$$
We now seek solutions to the deformation equations.
Let's make the ansatz
$$
{\delta}T\s=\Phi_B\s, \qquad {\delta}T_F\s=\Phi_F\s
\numbereq\name{\eqeleu}
$$
where $\Phi_F(\Phi_B)$ is the fermionic (bosonic) component of
a superfield of dimension $(d, \overline d)$.
Substituting into the deformation equations and using Eq. (\eqana) we
find that our ansatz of $\delta T$ and ${\delta}T_F$
satisfy the deformation
equations if the conformal dimension of $\Phi_F\s$ is
$d={\overline d}={1\over 2}$ and the dimension of $\Phi_B\s$
is $d=\overline d=1$.
These are the supersymmetric
generalizations of the so called {\it canonical deformations} found
in references \ref{\movr}{M. Evans and B. Ovrut, \pr41 3149  (1990),
M. Campbell, P. Nelson and E.
Wong, \imp6 4909 (1991).}.

For simplicity, consider a superconformal field theory of free scalars
and free two-dimensional fermions,
defined by the stress-tensor and the supercurrent
$$
\eqalign{
T\s&=-{1\over 2}{\eta^{\mu\nu}}{\partial}X_{\mu}
{\partial}X_{\nu}-{1\over 2}{\eta^{\mu\nu}}{\psi_{\mu}}{\partial}
{\psi_{\nu}} \cr
T_{F}\s&=-{1\over 2}{\eta}^{\mu\nu}{\psi_{\mu}}
{\partial}X_{\nu}. \cr}
\numbereq\name{\eqtoxou}
$$
The canonical deformations which correspond  to turning on a massless
NS-NS field, for example sending a weak gravitational and two-form
gauge wave through this background can be found by identifying
the bosonic component $\Phi_B(\sigma)$ of Eq. (\eqeleu)  with the
appropriate vertex operator ,
$$
{\delta}T\s=\Phi_B(\sigma)=K^{\mu\nu}(X)
{\overline{\partial}}X_{\nu}{\partial}X_{\mu}+{\partial^{\lambda}}
K^{\mu\nu}(X)
{\overline{\partial}}X_{\nu}{\psi_{\lambda}}{\psi_{\mu}}.
\numbereq\name{\eqmauro}
$$
The right hand side of this equation is a $(1, 1)$ primary field only if
the functions $K^{\mu\nu}(X)$ satisfy
$$
\Box K^{\mu\nu}(X)=0, \qquad \partial_\mu K^{\mu\nu}(X)=0.
\numbereq\name{\eqgewr}
$$
Its superpartner $\Phi_F(\sigma)$ is then found by calculating
the commutator of $\Phi_{B}\s$ with the supercurrent $T_{F}(\sigma)$.
We find that
$$
{\delta}T_F\s=K^{\mu\nu}(X){\overline{\partial}}X_{\nu}{\psi_{\mu}}.
\numbereq\name{\eqgeorg}
$$
By a tedious but rather straightforward calculation we can now verify
that $\delta T$ and $\delta T_F$ satisfy the deformation equations.
These superconformal deformations have also been found in papers \ref{\ovr}
{ B. Ovrut and S. Kalyana Rama, \pr45 550  (1992),
J. C. Lee, \zp54  283  (1992).}.

\newsection Spacetime Symmetries.

Let us recall the essential features of the approach to spacetime symmetries
which was developed in references [\movr],
\ref{\mg}{M. Evans and I. Giannakis, \pr44
2473 (1991).}.
Given any algebra of operators we can construct another algebra
isomorphic to the first one by means of a similarity transformation,
also called inner automorphism $\rho_h(O(\sigma))=e^{ih}O(\sigma)e^{-ih}$
or in infinitesimal form $\rho_h(O(\sigma))=O(\sigma)+i[h, O(\sigma)]$.
For any infinitesimal operator $h$ then the superconformal field
theories specified by $T_{\Phi}, T_{F(\Phi)}$ and
$T_{\Phi}+i[h, T_{\Phi}], T_{F(\Phi)}+i[h, T_{F(\Phi)}]$ are
isomorphic. Thus if 
$$
\eqalign{
{\delta}T&=T_{\Phi+{\delta{\Phi}}} -T_{\Phi}=i[h, T_{\Phi}] \cr
{\delta}T_F&=T_{F(\Phi+{\delta{\Phi}})} -T_{F(\Phi)}=i[h, T_{F(\Phi)}] \cr}
\numbereq\name{\eqlouc}
$$
then it follows that  $\Phi \mapsto {\Phi+{\delta{\Phi}}}$ is a symmetry
transformation of the spacetime fields.
As it should be obvious by now any inner automorphism preserves the
physics but not every automorphism can be interpreted as a change in
the spacetime fields. We should then think of Eq. (\eqlouc) as a restriction
on the operators $h$.

We now know how the stress-tensor $T$ and the supercurrent
$T_F$ deform as we change the spacetime
fields, Eq. (\eqeleu). We then must  find operators $h$ that when
commuted with $T$ and $T_F$ yield Eq. (\eqlouc).
Let $\Psi(\sigma)$ be a sum of superfields of dimension
$({1\over 2}, 0)$ and $(0, {1\over 2})$ and $h$ to be
$$
h={\int}d{\sigma}d{\theta}d{{\overline{\theta}}}{\Psi}(\sigma,
\theta, {\overline{\theta}}).
\numbereq\name{\eqkarapial}
$$
It then follows from the definition of superprimary fields that
$i[h, T]$ and $i[h, T_F]$ reproduces a ${\delta}T$ 
and ${\delta}T_F$of the form of  a canonical deformation Eq. (\eqeleu) and
thus can be interpeted as change in the spacetime fields.
The obvious choice for the operator $h$ is
$$
h={\ints}d{\theta}{\xi}^{\mu}({\chi})D{\chi}_{\mu}
=\ints\left(\xi^\mu(X)\dx_\mu+{\partial^{\mu}}{\xi^{\nu}}(X)
{\psi_{\mu}}{\psi_{\nu}}\right)(\sigma),
\numbereq\name{\eqantaz}
$$
where $\chi^\mu=\psi^{\mu}+{\theta}X^{\mu}$ and $D$ is the
covariant derivative.
This is the supersymmetric generalization of the
operator that generates coordinate and two form gauge transformations
about flat spacetime in
the bosonic string theory.
The integrand is only superprimary and of dimension $1\over 2$ if the
parameter $\xi$ satisfies
$$
\Box\xi^\mu(X)=0, \qquad \partial_\mu\xi^\mu(X)=0.
\numbereq\name{\eqivic}
$$
These conditions are required because of normal ordering.
We then proceed to calculate the commutator of $h$ with $T$ and
$T_F$ and we find
$$
\eqalign{
i[h, T\s]&={\partial^{\mu}}{\xi^{\nu}}
{\overline{\partial}}X_{\nu}{\partial}X_{\mu}+{\partial^{\lambda}}
{\partial^{\mu}}{\xi^{\nu}}
{\overline{\partial}}X_{\nu}{\psi_{\lambda}}{\psi_{\mu}} \cr
i[h, T_F\s]&={1\over 2} {\partial^{\mu}}{\xi^{\nu}}
{\overline{\partial}}X_{\nu}{\psi_{\mu}}. \cr}
\numbereq\name{\eqgokic}
$$
The result provides us with the transformation properties
of the physical fields under coordinate and two form gauge
transformations
$$
\delta K^{\mu\nu}=\partial^{\mu}{\xi^{\nu}}.
\numbereq\name{\eqofori}
$$
It is not hard to generalize this construction to an infinite class of
infinitesimal gauge symmetries \ref{\mgn}{M. Evans, I. Giannakis and
D. Nanopoulos \pr50 4022 (1994).}
and to finite
symmetry transformations (T-duality) \ref{\gm}{M. Evans and I. Giannakis,
\np472  139  (1996),
I. Giannakis, \pl388  543 (1996).}.
It is also
worth noting that this approach treats on equal footing
exact and spontaneously broken spacetime symmetries
\ref{\bg}{J. Bagger and I. Giannakis, \pr56 2317 (1997).}.

Canonical deformations have a number of interesting features:
superprimary fields of dimension $({1\over 2}, {1\over 2})$ are
in natural correspondence with the physical states of string
theory, being the vertex operators. As such they have a nice
spacetime interpretation in terms of turning on spacetime fields.
Appealing though they are canonical deformations have also
significant drawbacks. They correspond to turning on
NS-NS fields in a particular gauge as we have seen while they
do not appear to describe spacetime fermions and R-R bosonic
fields which are described in terms of spin fields. Spin fields cannot be
written as superfields. These string backgrounds have
attracted interest recently due to the conjectured AdS/CFT
equivalence \ref{\mal}{J. Maldacena, \amp2 231 (1998),
S. S. Gubser, I. R. Klebanov and A. M. Polyakov, \pl428
105 (1998), E. Witten, \amp2 253 (1998)}.
We might attempt to identify
the bosonic component $\Phi_{B}(\sigma)$ of the
canonical deformation with the appropriate spacetime fermionic
vertex operator
$$
\delta T(\sigma)=\Phi_{B}(\sigma)=
{\Psi}_{\mu}^{\alpha}(X)S_{\alpha}
e^{-{{\phi}\over 2}}{\overline{\partial}}X^{\mu}
\numbereq\name{\eqaboua}
$$
This is a $(1, 1)$ primary field only if the functions ${\Psi}_{\mu}^{\alpha}(X)$
obey
$$
\Box  {\Psi}_{\mu}^{\alpha} (X)=0, \quad {\gamma}^{\mu}{\partial_{\mu}}
{\Psi}_{\nu}^{\alpha} (X)=0, \quad {\partial^{\mu}}
{\Psi}_{\mu}^{\alpha} (X)=0. 
\numbereq\name{\eqamana}
$$
In order then to find its superpartner we need to calculate the commutator
of $\Phi_{B}(\sigma)$ with the supercurrent $T_F(\sigma)$.
The commutator of the vertex operator which is
written in terms of spin fields with
the supercurrent $T_F$ is not well-defined since the corresponding
OPE in the complex plane involves branch cut singularities
$$
T_{F}(z){\Phi_B}(w)={{\gamma^{\lambda}}_{\alpha{\dot{\beta}}}
{\partial_{\lambda}}{\Psi}_{\mu}^{\alpha}(X)S^{\dot\beta}
e^{-{{\phi}\over 2}}{\overline{\partial}}X^{\mu} \over {(z-w)^{3\over 2}}}
+{{\gamma^{\lambda}}_{\alpha{\dot{\beta}}}
{\Psi}_{\mu}^{\alpha}(X)S^{\dot\beta}
e^{-{{\phi}\over 2}}{\partial}X_{\lambda}
{\overline{\partial}}X^{\mu} \over {(z-w)^{1\over 2}}}
\numbereq\name{\eqtzenam}
$$
This suggests then that the
deformations we have just constructed in terms of superfields
are not the most general solution to the deformation equations.
In a forthcoming publication we intend to present the resolution
to these problems.

\newsection Acknowledgments.
This work was done in collaboration with Jonathan Bagger and Mark Evans.
I would like to
thank James T. Liu and M. Porrati for useful discussions. This work was
supported by NSF grant PHY-9722083.
 
\immediate\closeout1
\bigbreak\bigskip

\line{\twelvebf References. \hfil}
\nobreak\medskip\vskip\parskip

\input refs

\vfil\end

\bye